\documentclass[a4paper,fleqn]{article}
\usepackage{a4,graphicx,cite}

\begin{document}

\title{\bf\Large Quasiparticle Calculations for Point Defects on Semiconductor Surfaces}

\author{\sc\normalsize Magnus Hedstr\"om\thanks{Corresponding author; e-mail: hedstrom@fhi-berlin.mpg.de}\ ,
Arno Schindlmayr, and
Matthias Scheffler\\[1ex]
\it\normalsize Fritz-Haber-Institut der Max-Planck-Gesellschaft, Faradayweg 4--6,\\
\it\normalsize 14195 Berlin-Dahlem, Germany}
\date{\normalsize (August 19, 2002)}

\maketitle

\begin{abstract}
We discuss the implementation of quasiparticle calculations for point
defects on semiconductor surfaces and, as a specific example,
present an ab initio study of the electronic structure
of the As vacancy in the +1 charge state on the GaAs(110)
surface. The structural properties are calculated with the
plane-wave pseudopotential method, and the quasiparticle
energies are obtained from Hedin's $GW$ approximation.
Our calculations show that the 1a$''$ vacancy state in the band gap
is shifted from 0.06 to 0.65 eV above the valence-band maximum
after the self-energy correction to the Kohn-Sham eigenvalues. The $GW$ result
is in close agreement with a recent surface photovoltage imaging measurement.\\[1ex]


\end{abstract}

\section{Introduction\label{sec:introduction}}

Over the past 15 to 20 years intense research efforts have been made to
understand the nature of native point defects near and at
semiconductor surfaces. Such defects play an important role in surface
electrical characteristics, e.g., Fermi-level pinning and charge-carrier
recombination. They also influence the kinetics of adsorption, diffusion,
and growth as well as surface chemical activity. The progress of the
research efforts in understanding point defects on III--V semiconductor
surfaces has recently been reviewed by Ebert \cite{ebert2001}.

At the atomistic level significant insight into the identification of
these defects comes from scanning tunneling microscopy (STM).
However, STM measures the local electronic density of states, which
cannot always be directly interpreted as the atomic geometry.
Therefore, accurate calculations have turned out to be very valuable
for the interpretation of the experimental findings.
In this paper we present an ab initio study of the
arsenic vacancy $V_{\rm As}$ on the GaAs(110) surface
under p-type conditions.
In STM images of the filled-state As sublattice, this vacancy gives
rise to two distinct features: a dark hole of the lateral dimension
of one As dangling bond and a long-range charge-related
depression of the neighbouring As atoms \cite{lengel94}.
The first feature is directly related to the missing atom, and
the second one is due to a downward local band bending, from which
it was concluded that in p-GaAs the arsenic vacancy
is positively charged.
STM images acquired under positive bias probe the empty
p$_{z}$ orbitals of the Ga sublattice and show an
enhancement of the contrast from the two Ga atoms nearest to
the vacancy. Lengel et al.\ \cite{lengel94} interpreted their
finding as an outward relaxation of the two Ga atoms neighbouring
the vacancy. They also found support for this interpretation from
a tight-binding calculation. Their calculation suggested
a charge state of +2 for the vacancy. On the other hand,
by comparing STM images of the arsenic vacancy with the images
of Zn dopant atoms on p-type GaAs(110), Chao et al.\
\cite{chao96} were
able to determine the charge state of  $V_{\rm As}$
to be +1, utilizing the fact that the dopant atom is in a charge
state of $-1$ a priori.

Recent density-functional theory (DFT) calculations by Zhang and
Zunger \cite{zhang96} and by Kim and Chelikowsky \cite{kim96,kim98}
have confirmed the stability of the +1 charge state. Furthermore,
their calculations have independently shown that the energy minimization
of the defect geometry results in a downward relaxation of the
neighbouring Ga atoms, and the calculated STM images show
an enhancement of these two atoms in agreement with experiment.
Both of these calculations were performed at a comparable level of
sophistication but differed in the details.
Zhang and Zunger \cite{zhang96} found an asymmetric relaxation of
$V^{+}_{\rm As}$, whereas Kim and Chelikowsky \cite{kim96,kim98}
found a symmetric relaxation, which seems to agree with
experiment. However, in a recent combined experimental and
theoretical study on a related system, the phosphorus vacancy on
InP(110), Ebert et al.\ \cite{ebert2000} suggested that the symmetric
feature of the experimental STM image could also be explained by a
thermal flip motion from two degenerate asymmetric configurations.

For properties other than structural ones the experimental situation
is less complete. There are so far no direct measurements of
either defect ionization or charge-transfer levels for
$V_{\rm As}$ on GaAs(110). Indirect information about these quantities can
be obtained, for instance, from measurements of the local band bending.
Both the ionization and charge-transfer levels can be calculated
theoretically, but so far this has not been done beyond DFT and
using the local-density approximation (LDA). Furthermore, we note
that the reported values deviate among the different studies.
DFT-LDA also suffers from the underestimation of the fundamental
band gap, which introduces
a large uncertainty when interpreting the Kohn-Sham eigenvalues
of defect levels that lie inside the band gap as actual energy
levels. As we will show in this paper, the band-gap problem for
defect states can be circumvented by employing the $GW$
approximation \cite{hedin65} for the electronic self-energy and
calculating the proper
quasiparticle energies. The $GW$ approximation has previously
been applied to the perfect GaAs(110) surface \cite{zhu89,pulci98}
as well as to bulk defects in some other materials, such as the Li
vacancy in LiCl \cite{surh95} and the oxygen vacancy in zirconia
\cite{kralik98}, with good agreement between the theoretical
predictions and the available experimental data. We note in
passing that all the mentioned studies relied on plasmon-pole
models, whereas our study takes the full frequency dependence
of the screened Coulomb interaction into account.

The paper is organized as follows. The method and computational
details are described in Section \ref{sec:method} and
results from the calculations in Section \ref{sec:results}.
Finally, Section \ref{sec:conclusions} summarizes our conclusions.

\section{Method and Computational Details\label{sec:method}}

\subsection{Atomic geometry}

The ground-state properties, such as the geometric structure of the defect,
are calculated using DFT \cite{hohenberg64,kohn65} and the plane-wave
pseudopotential method as implemented in the FHImd code \cite{bockstedte97,fuchs99}.
The exchange-correlation functional is treated in the local-density approximation
as parametrized by Perdew and Zunger \cite{perdew81}.
The single-particle orbitals are expanded in plane waves with a
cutoff energy of 10 Ry in most of the calculations, but convergence
tests were performed up to a cutoff of 15 Ry.
With the basis set used we find the bulk lattice constant to be 5.55 \AA\
(neglecting zero-point vibrations), which is in agreement with other
calculations for III--V semiconductors \cite{kim98,alves91,schwarz98}
and 1.8\% smaller than the experimental value at room temperature
of 5.65 \AA\ \cite{adachi85}.

\begin{figure}
\includegraphics[width=0.9\textwidth]{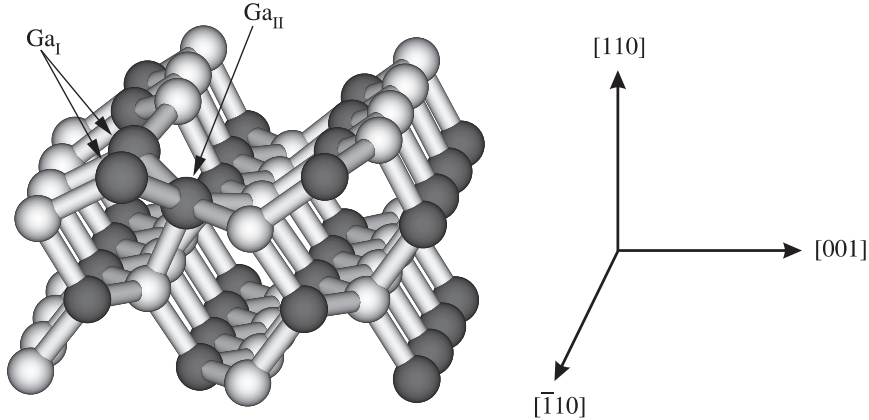}
\caption{\label{fig:fig1} 
A 4$\times$2 surface unit cell with an As vacancy. The As atoms are 
shown in light grey and the Ga atoms in dark grey. The three Ga atoms
next to the vacancy are indicated by arrows. Note the five-fold
coordination of the Ga$_{\rm II}$ in the second layer.}
\end{figure}

The As vacancy on GaAs(110), whose geometry is shown in Fig.\ \ref{fig:fig1},
is described using the supercell method
with periodic boundary conditions. The supercell slab consists of
a 4$\times$2 surface unit cell with six atomic layers, with one As
atom missing in the top layer, and four vacuum layers.
The dangling bonds at the bottom of the slab are
passivated by pseudoatoms with noninteger nuclear charges of
0.75 and 1.25 for As and Ga termination, respectively.
We allow the atoms in the top three layers to fully relax while
keeping the atoms in the three bottom layers at their theoretical 
bulk positions. In the case of the positively charged defect we
apply a uniform compensating charge throughout the unit cell
in order to maintain overall charge neutrality. The integration in reciprocal
space was done using the special ${\bf k}$-point
$(\frac{1}{4},\frac{1}{4},0)$ \cite{baldereschi73} in the irreducible
part of the Brillouin zone.

In order to test the influence of the supercell size
we also performed a series of investigations using smaller 2$\times$2
and 2$\times$1 unit cells and found that the dispersion of the defect
level, due to the spurious interaction of the defect with its periodic images,
changed from 0.6 eV (2$\times$2) to 1.5 eV (2$\times$1), compared
to 0.3 eV in the case of the 4$\times$2 unit cell.
For the 2$\times$1 unit cell the defect--defect interaction is so
strong that the two defect levels in the band gap exhibit a crossing,
thus making any further calculations using such a small unit cell
doubtful. On the other hand, the 2$\times$2 unit cell gave results
consistent with the ones obtained for the larger 4$\times$2 cell.
In particular, the formation energy for the neutral defect
turned out to be the  same using either of the unit cells.

Furthermore,
we did a series of comparisons between six-layer slabs and four-layer
slabs for the 2$\times$2 unit cell and found no significant
differences in the defect band structures (dispersion relations) for any
of the possible charge states +1, 0, or $-1$. In the case of the
four-layer slab only the top two layers were relaxed in the geometry
optimization.

\subsection{Quasiparticle energies}

Green-function theory is a mathematical framework for calculating
the actual quasiparticle band structure, i.e., electron addition and
removal energies, which characterize the excitation from an
$N$-particle system to an ($N$$\pm$1)-particle system during
a photoemission process. These energies,
$\epsilon_{n{\bf k}}^{\rm qp}$, are in principle given exactly
by the quasiparticle equation
\begin{equation}\label{eq:quasi}\hspace{-\mathindent}
\left( -\frac{1}{2} \nabla^2 + V_{\rm ext}({\bf r}) +V_{\rm H}({\bf r}) \right)
\psi_{n{\bf k}}^{\rm qp}({\bf r})
+ \int \Sigma({\bf r},{\bf r}';\epsilon_{n{\bf k}}^{\rm qp})
\psi_{n{\bf k}}^{\rm qp}({\bf r}') \,d^{3}r'
= \epsilon_{n{\bf k}}^{\rm qp} \psi_{n{\bf k}}^{\rm qp}({\bf r}) \;,
\end{equation}
where $V_{\rm ext}$ is the external potential,
$V_{\rm H}$ the Hartree potential, and
$\psi_{n{\bf k}}^{\rm qp}$
the quasiparticle wave function.
$\Sigma$ denotes the self-energy operator, which contains all contributions
from dynamic exchange and correlation. In general the self-energy is
nonlocal, energy-dependent, and has a nonzero imaginary part, whose
inverse is proportional to the lifetime of the excited state. For real systems
$\Sigma$ cannot be calculated exactly and must be approximated by a suitable
functional expression. In the $GW$ approximation $\Sigma$ is
given by \cite{hedin65}
\begin{equation}\label{eq:sigma}
\Sigma({\bf r},{\bf r}';i\tau) = i G({\bf r},{\bf r}';i\tau) W({\bf r},{\bf r}';i\tau) \;,
\end{equation}
where $G$ is the one-particle Green function and
$W$ the dynamically screened Coulomb interaction. All operators
are here written in real space and imaginary time. This representation
has the advantage that the self-energy is a simple
product, which is exploited in the $GW$ space-time method
\cite{rieger99,steinbeck2000} that we have used for our
calculations. It avoids the multidimensional convolutions in
reciprocal space and on the frequency axis that must be
evaluated in conventional implementations.
The inclusion of dynamic screening in the self-energy describes
the correlation hole around individual electrons due to
the Coulomb interaction. In this way the $GW$ approximation
goes beyond Hartree-Fock theory, which only describes exchange effects
and ignores the correlation of electrons with different
spin. Instead of
simplified plasmon-pole models, which have been employed in almost
all $GW$ calculations for surfaces so far, we use the
random-phase approximation with the full frequency dependence
for the screened Coulomb interaction. At the end of the calculation
the self-energy is Fourier transformed and analytically continued
to the real energy axis.

As usual, also in our calculations the Green function is
obtained from the Kohn-Sham eigenfunctions, incorporating a
large number of unoccupied states. The LDA
wave functions form a reasonable starting point for the study of
electronic band structures and are obtained from an equation closely
resembling Eq.\ (\ref{eq:quasi}), with the local exchange-correlation
potential, $V_{\rm xc}$, in the place of the self-energy.
The quasiparticle energies are therefore
calculated from first-order perturbation theory according to
\begin{equation}\label{eq:qp}
\epsilon_{n{\bf k}}^{\rm qp} = \epsilon_{n{\bf k}}^{\rm LDA} +
\langle \psi_{n{\bf k}}^{\rm LDA} | \Sigma (\epsilon_{n{\bf k}}^{\rm qp})
- V_{\rm xc} | \psi_{n{\bf k}}^{\rm LDA} \rangle \;,
\end{equation}
where $\epsilon_{n{\bf k}}^{\rm LDA}$ is the corresponding LDA eigenvalue
associated with the wave function $\psi_{n{\bf k}}^{\rm LDA}$.

\begin{figure}
\includegraphics[height=7cm]{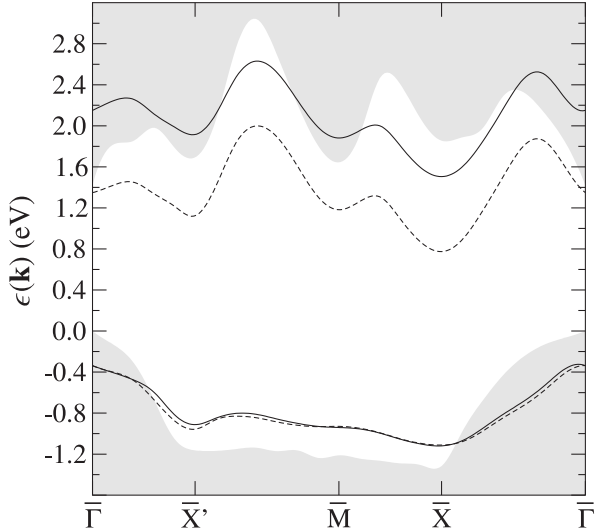}
\caption{\label{fig:fig2}
Calculated surface band structure
of relaxed GaAs(110): dashed lines, DFT-LDA; solid lines, $GW$.
The grey-shaded regions show the projected $GW$ bulk bands.}
\end{figure}

Before investigating the $V^{+}_{\rm As}$ system
we first perform accurate convergence
tests for the perfect surface, which can be represented
with a smaller 1$\times$1 surface unit cell. Fig.\ \ref{fig:fig2}
shows the quasiparticle band structure for the GaAs(110) surface.
The dashed lines are obtained from  DFT-LDA with a cutoff energy
of 15 Ry, and the solid lines are the $GW$ quasiparticle energies.
The projected $GW$ bulk band structure is indicated in grey.
The ground-state density used for the DFT band-structure calculation
was obtained using
four special Monkhorst-Pack ${\bf k}$-points \cite{monkhorst76}, which
ensures the same Brioullin-zone sampling as a single ${\bf k}$-point
for the larger 2$\times$2 unit cell.
The self-energy operator was evaluated at the
four high-symmetry ${\bf k}$-points ($\bar{\Gamma}$,\={X},\={M},\={X}').
The LDA surface bands are matched to the projected bulk bands
by aligning the electrostatic potential in
the central region of the surface slab with that of the bulk.
The $GW$ approximation has only a small effect on the dispersion of the
occupied surface band and does not change its relative distance to the
valence-band edge at $\bar{\Gamma}$. The $GW$ and LDA occupied
surface bands are hence aligned at this point. From Fig.\ \ref{fig:fig2} we
then observe an almost uniform upward shift of the $GW$ corrected
unoccupied surface band by 0.8 eV, which is in close agreement with Zhu
et al.\ \cite{zhu89}
and 0.1--0.2 eV less than found by Pulci et al.\ \cite{pulci98}.
This opening of the gap makes for
the typical improvement of $GW$ quasiparticle band
structures over DFT-LDA for the unoccupied states. We also note that the
shift in the surface band is similar to the shift we find for the
conduction bands in the bulk, 0.7 eV, illustrating the similarity of the
wave-function character between the surface states and the
bulk valence and conduction bands. The filled surface state at and in
the vicinity of the $\bar{\Gamma}$-point appears as a surface
resonance, i.e., hybridization with extended bulk states is
noticeable, and the peak becomes broad.
The band structure at this point is shown for the
sole purpose of illustrating the LDA and $GW$ alignment.
We repeated the calculations using different cutoff energies and found
that the $GW$ shift of 0.8 eV remained for 13, 11, and 10 Ry cutoff.
Also, for a four-layer slab we obtained a $GW$ correction of
0.8 eV, and at 10 Ry cutoff we investigated the sensitivity of the
$GW$ corrected band structure to variations in the number of unoccupied
bands. The self-energy is fully converged if 1049 unoccupied bands are
included. The difference in the quasiparticle energies
around the band gap is less
than 0.05 eV when the number of unoccupied bands is reduced to
153. With 379 unoccupied bands the agreement is better than 0.02 eV,
which suffices for our purpose.
For the $GW$ calculation of the defect levels in the 2$\times$2
surface unit cell we hence use 1500 unoccupied
bands, which correspond to 379 bands in the case of a
1$\times$1 cell.

\section{Results and Discussion\label{sec:results}}

\subsection{Structural properties}

Let us first recall the structure of the perfect GaAs(110) surface,
which corresponds to the defect-free right-hand part of Fig.\ \ref{fig:fig1}.
This surface belongs to the C$_{1{\rm h}}$ point group, which consists of a
single mirror plane perpendicular to the [\={1}10] direction.
The relaxation of the
surface preserves the point-group symmetry and consists chiefly of
an inward movement of the Ga atom and an outward movement of the As
atom. Hence the relaxation can be characterized mainly as a pure bond
rotation, which results in a buckling of the surface \cite{alves91}.
The Ga atoms rehybridize from an sp$^{3}$ to an sp$^{2}$ bonding situation to
form a locally planar structure. The empty p$_{z}$-like
orbital perpendicular to this plane forms the unoccupied
surface band. The As atom maintains its sp$^{3}$ hybridization
and binds to three Ga atoms in a pyramidal arrangement with
a nonbonding electron pair in the fourth direction
pointing away from the surface. The filled surface band is composed
from the As lone electron pairs.

We now turn the discussion to the As vacancy. The removal of the
As atom leaves a dangling bond on each of the three Ga atoms surrounding
the vacancy. As mentioned earlier, there is conflicting evidence for
\cite{zhang96,ebert2000} and against \cite{kim96,kim98} a breaking of the
mirror symmetry due to the relaxation of the positively charged
anion vacancy, which requires further careful studies.
In this work we choose the second scenario and perform a relaxation
of $V^{+}_{\rm As}$ with the constraint that the C$_{1{\rm h}}$ symmetry
is preserved. The calculated equilibrium geometry shows a downward
relaxation of the two Ga atoms in the first layer (Ga$_{\rm I}$)
and an upward movement of the second-layer Ga atom (Ga$_{\rm II}$), which
forms two new weak bonds with the Ga$_{\rm I}$ atoms. Thus the
coordination of Ga$_{\rm II}$ is increased to five. In Fig.\
\ref{fig:fig1} the Ga$_{\rm I}$ and Ga$_{\rm II}$ atoms are indicated by arrows.
We find the bond length
between Ga$_{\rm I}$ and Ga$_{\rm II}$ to be 2.89 \AA\ for $V^{+}_{\rm As}$,
which is slightly longer than the value of 2.78 \AA\ reported in Ref.\ \cite{kim96}.
When going from $V^{+}_{\rm As}$ to $V^{-}_{\rm As}$,
the Ga$_{\rm I}$--Ga$_{\rm II}$ bond length decreases, indicating the bonding
character of the 1a$''$ defect level.

\subsection{Electronic properties}

\begin{figure}
\includegraphics[height=7cm]{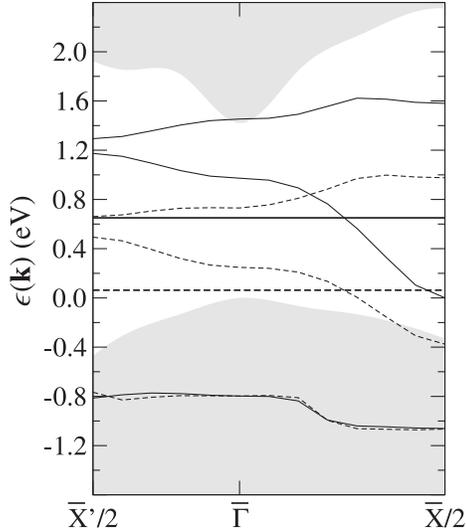}
\caption{\label{fig:fig3}
Calculated band structure
of $V^{+}_{\rm As}$ on GaAs(110): dashed lines, DFT-LDA; solid lines, $GW$.
The horizontal lines mark the actual 1a$''$ defect level as obtained at the
special ${\bf k}$-point.}
\end{figure}

The arsenic vacancy gives rise to three electronic states, 1a$'$, 1a$''$,
and 2a$'$, where a$'$ denotes states that are even
with respect to the mirror plane and a$''$ denotes a state that is odd.
The 1a$'$ state is located several eV below the valence-band maximum
and is thus always filled \cite{kim98}. The 1a$''$ state is located in
the Kohn-Sham band gap, and depending on the location of the Fermi level, it is
either empty or filled with one or two electrons.
The 1a$''$ state has a nodal plane coinciding with the mirror plane
but has bonding character between Ga$_{\rm I}$ and Ga$_{\rm II}$ as mentioned
above. The 2a$'$ state is found close to the conduction-band minimum (CBM)
in the case of $V^{+}_{\rm As}$ and clearly above the CBM
for $V^{0}_{\rm As}$ and  $V^{-}_{\rm As}$.
Since the Kohn-Sham approach underestimates the band gap, it is plausible that
the energies of the 1a$''$ and 2a$'$ vacancy levels are incorrectly
given by DFT-LDA. We will focus on the 1a$''$ level for
$V^{+}_{\rm As}$. This state is mainly located at the
Ga$_{\rm I}$--Ga$_{\rm II}$ bonds and thus has a character different
from any other in either the surface or the bulk, where there are
no Ga--Ga bonds present. Therefore, one cannot a priori predict the
quasiparticle correction from the shift of the normal
surface or bulk bands;
it has to be calculated separately.
Our $GW$ calculations for the shift of the defect level are done
for a four-layer 2$\times$2 unit cell, using the convergence parameters
that we found satisfactory for the clean surface.
The defect--defect interaction gives rise to a rather large dispersion and
requires a careful treatment in order
to extract the most accurate value for the defect level. For this reason we
fit the calculated DFT-LDA defect band to a simple
tight-binding (TB) model.
Since the 1a$''$ state is odd with respect to the mirror plane, one can,
for the purpose of TB, regard it as an orbital with p symmetry
that forms $\pi$-type bonds in the [001] direction
and exhibits $\sigma$-bonding along the [\={1}10] direction.
Within our supercell approach the defect sites form a
rectangular lattice with lattice parameters $a_{x}$ and $a_{y}$,
where $a_{x}$ is the length of the supercell along [\={1}10]
and $a_{y}$ the length along [001]. We found that we could fit
the dispersion of the DFT-LDA (and also the $GW$) defect band by
\begin{eqnarray}
\epsilon({\bf k}) &=& \epsilon_{\rm d} +
2 V_{1\sigma} \cos(k_{x}a_{x}) + 2 V_{1\pi} \cos(k_{y}a_{y}) +
V_{2} \cos(k_{x}a_{x}) \cos (k_{y}a_{y})\nonumber\\
&&\mbox{} + 2 V_{3\sigma} \cos(2k_{x}a_{x}) + 2 V_{3\pi} \cos(2k_{y}a_{y}) \;,\label{eq:TB}
\end{eqnarray}
where the fitting parameters $\epsilon_{\rm d}$
and $V_{n\eta}$ have the meaning
of the energy of a single defect and the hopping integrals,
respectively. In order to properly reproduce the Kohn-Sham dispersion we
included interactions up to third-nearest neighbours ($V_{3\sigma}$
and $V_{3\pi}$). The correlation between the DFT-LDA results and the TB fit
turned out to be 0.9999. At the special ${\bf k}$-point
${\bf k}_{\rm s}=\frac{2\pi}{4}(a_{x}^{-1},a_{y}^{-1},0)$
the contribution from the first and second neighbours vanishes
and $\epsilon ({\bf k}_{\rm s}) = \epsilon_{\rm d}
+ 2 V_{3\sigma} \cos(2k_{x}a_{x}) + 2 V_{3\pi} \cos(2k_{y}a_{y})$.
As the contribution from the third
neighbours is very small, 0.06 eV, we feel satisfied to directly take
the DFT-LDA calculated value at ${\bf k}_{\rm s}$ as the defect level.
Furthermore, the TB fit to the $GW$ dispersion gives a similar
estimate of $V_{3\sigma}$ and $V_{3\pi}$ as in DFT-LDA,
so when we calculate the quasiparticle
correction to the defect level the errors largely cancel.
The results for $V^{+}_{\rm As}$ are shown in Fig.\ \ref{fig:fig3},
where the DFT-LDA results are marked with dashed lines and the $GW$
results with solid lines. The $GW$ and the DFT-LDA results are aligned at
${\bf k}_{\rm s}$ using the occupied surface band derived from
the As electron lone pairs similarly as was done for the perfect surface.
The As atoms contributing to this band are all situated in the
defect-free row of surface As atoms along [\={1}10]. In Fig.\
\ref{fig:fig3} this state appears in the projected bulk bands (grey shaded)
at $-0.8$ eV. From our calculation we find the quasiparticle correction to
the 1a$''$ state to be 0.59 eV. However, the four-layer slab is
slightly too thin to allow an accurate alignment of the Kohn-Sham band
structure to the projected bulk bands, so for this purpose we use a six-layer
4$\times$2 unit cell, for which we also expect the
agreement between the actual defect level and the value
at ${\bf k}_{\rm s}$ to be even better than in the case of
the 2$\times$2 cell. At the DFT-LDA level the defect state is
found to be 0.06 eV above the valence-band maximum and is indicated by
a dashed horizontal line in Fig.\ \ref{fig:fig3}. The $GW$ correction
adjusts this upward to 0.65 eV (solid horizontal line). The calculated
values for the 2a$'$ state are also shown in the figure.

So far no direct photoemission measurement for this defect level
has been reported in the literature. Hence we compare our results to a
measurement by Aloni et al.\ \cite{aloni99,aloni2001}, who used
atomically resolved surface photovoltage imaging with STM.
They measured the tip-induced band bending by scanning along the [001]
direction of the surface and found a pinning of the band bending at 0.53 eV,
at the location of a single defect. When the tip bias was changed, the
band bending remained the same at the defect site until the tip-induced
band bending in the defect-free region exceeded 0.53 eV. At further
negative tip-sample bias the signature of the defect disappeared,
and the band bending was the same at the defect site as in the defect-free
regions, from which one can conclude that at this point the tip-induced band
bending had pushed the defect state clearly below the Fermi level,
which changes the defect into the neutral charge state.
From the knowledge of the bulk Fermi level the authors concluded
that the defect level is located at 0.62$\pm$0.03 eV above the
valence-band maximum for a flat-band situation. Our result of 0.65 eV
turns out to be remarkably close to the experimental value, although it is
not entirely clear to which extent the latter also includes energy contributions
from atomic relaxation processes. We estimate that these are small, however,
because our DFT-LDA total-energy calculations for the $V^{+}_{\rm As}$ and
$V^{0}_{\rm As}$ geometries, both with one electron occupying the defect
level, indicate a maximum energy gain of only 0.16 eV for a complete relaxation.
In any case, both the experiment and our
$GW$ calculation agree that the defect level is
closer to mid-gap than to the valence-band maximum.

\section{Conclusions\label{sec:conclusions}}

We have found the $GW$ approximation to be a useful tool for
investigating the electronic structure of point defects on
semiconductor surfaces. The methodology and the possible difficulties are
described in this paper. As an example we have studied the arsenic
vacancy on GaAs(110) in its positive charge state. While the LDA
underestimates the fundamental band gap and places the unoccupied
1a$''$ defect state just 0.06 eV above the valence-band maximum, the
$GW$ approximation opens the gap and shifts the defect level upward
to 0.65 eV. This theoretical result is in good agreement with a
recently reported experimental value of 0.62 eV that was
deduced from atomically resolved surface photovoltage imaging with STM.

\section*{Acknowledgements}

This work was funded in part by the EU through the NANOPHASE
Research Training Network (Contract No.\ HPRN-CT-2000-00167).
We thank J\"org Neugebauer and G\"unther Schwarz for helpful discussions.


\begin{thebibliography}{26}
\renewcommand{\itemsep}{-\parsep}
\bibitem{ebert2001} {\sc Ph. Ebert}, Curr. Opin. Solid State Mater. Sci. {\bf 5}, 211 (2001).
\bibitem{lengel94} {\sc G. Lengel}, {\sc R. Wilkins}, {\sc G. Brown}, {\sc M. Weimer}, {\sc J. Gryko}, and {\sc R. E. Allen}, Phys. Rev. Lett. {\bf 72}, 836 (1994).
\bibitem{chao96} {\sc K.-J. Chao}, {\sc A. R. Smith}, and {\sc C.-K. Shih}, Phys. Rev. B {\bf 53}, 6935 (1996).
\bibitem{zhang96} {\sc S. B. Zhang} and {\sc A. Zunger}, Phys. Rev. Lett. {\bf 77}, 119 (1996).
\bibitem{kim96} {\sc H. Kim} and {\sc J. R. Chelikowsky}, Phys. Rev. Lett. {\bf 77}, 1063 (1996).
\bibitem{kim98} {\sc H. Kim} and {\sc J. R. Chelikowsky}, Surf. Sci. {\bf 409}, 435 (1998).
\bibitem{ebert2000} {\sc Ph. Ebert}, {\sc K. Urban}, {\sc L. Aballe}, {\sc C. H. Chen}, {\sc K. Horn}, {\sc G. Schwarz}, {\sc J. Neugebauer}, and {\sc M. Scheffler}, Phys. Rev. Lett. {\bf 84}, 5816 (2000).
\bibitem{hedin65} {\sc L. Hedin}, Phys. Rev. {\bf 139}, A796 (1965).
\bibitem{zhu89} {\sc X. Zhu}, {\sc S. B. Zhang}, {\sc S. G. Louie}, and {\sc M. L. Cohen}, Phys. Rev. Lett. {\bf 63}, 2112 (1989).
\bibitem{pulci98} {\sc O. Pulci}, {\sc G. Onida}, {\sc R. Del Sole}, and {\sc L. Reining}, Phys. Rev. Lett. {\bf 81}, 5374 (1998).
\bibitem{surh95} {\sc M. P. Surh}, {\sc H. Chacham}, and {\sc S. G. Louie}, Phys. Rev. B {\bf 51}, 7464 (1995).
\bibitem{kralik98} {\sc B. Kr\'alik}, {\sc E. K. Chang}, and {\sc S. G. Louie}, Phys. Rev. B {\bf 57}, 7027 (1998).
\bibitem{hohenberg64} {\sc P. Hohenberg} and {\sc W. Kohn}, Phys. Rev. {\bf 136}, B864 (1964).
\bibitem{kohn65} {\sc W. Kohn} and {\sc L. J. Sham}, Phys. Rev. {\bf 140}, A1133 (1965).
\bibitem{bockstedte97} {\sc M. Bockstedte}, {\sc A. Kley}, {\sc J. Neugebauer}, and {\sc M. Scheffler}, Comp. Phys. Commun. {\bf 107}, 187 (1997).
\bibitem{fuchs99} {\sc M. Fuchs} and {\sc M. Scheffler}, Comp. Phys. Commun. {\bf 119}, 67 (1999).
\bibitem{perdew81} {\sc J. P. Perdew} and {\sc A. Zunger}, Phys. Rev. B {\bf 23}, 5048 (1981).
\bibitem{alves91} {\sc J. L. A. Alves}, {\sc J. Hebenstreit}, and {\sc M. Scheffler}, Phys. Rev. B {\bf 44}, 6188 (1991).
\bibitem{schwarz98} {\sc G. Schwarz}, {\sc A. Kley}, {\sc J. Neugebauer}, and {\sc M. Scheffler}, Phys. Rev. B {\bf 58}, 1392 (1998).
\bibitem{adachi85} {\sc S. Adachi}, J. Appl. Phys. {\bf 58}, R1 (1985).
\bibitem{baldereschi73} {\sc A. Baldereschi}, Phys. Rev. B {\bf 7}, 5212 (1973).
\bibitem{rieger99} {\sc M. M. Rieger}, {\sc L. Steinbeck}, {\sc I. D. White}, {\sc H. N. Rojas}, and {\sc R. W. Godby}, Comp. Phys. Commun. {\bf 117}, 211 (1999).
\bibitem{steinbeck2000} {\sc L. Steinbeck}, {\sc A. Rubio}, {\sc L. Reining}, {\sc M. Torrent}, {\sc I. D. White}, and {\sc R. W. Godby}, Comp. Phys. Commun. {\bf 125}, 105 (2000).
\bibitem{monkhorst76} {\sc H. J. Monkhorst} and {\sc J. D. Pack}, Phys. Rev. B {\bf 13}, 5188 (1976).
\bibitem{aloni99} {\sc S. Aloni}, {\sc I. Nevo}, and {\sc G. Haase}, Phys. Rev. B {\bf 60}, R2165 (1999).
\bibitem{aloni2001} {\sc S. Aloni}, {\sc I. Nevo}, and {\sc G. Haase}, J. Chem. Phys. {\bf 115}, 1875 (2001).
\end{thebibliography}
\end{document}